# High Performance Implementation of Boris Particle Pusher on DPC++. A First Look at oneAPI[*]

Valentin Volokitin[1], Alexey Bashinov[2], Evgeny Efimenko[2], Arkady Gonoskov[3], Iosif Meyerov[1,4]

[1] Mathematical Center, Lobachevsky University, Nizhni Novgorod 603950, Russia
[2] Institute of Applied Physics, Russian Academy of Sciences, Nizhny Novgorod 603950, Russia
[3] Department of Physics, University of Gothenburg, SE-41296 Gothenburg, Sweden
[4] `meerov@vmk.unn.ru`

**Abstract.** New hardware architectures open up immense opportunities for supercomputer simulations. However, programming techniques for different architectures vary significantly, which leads to the necessity of developing and supporting multiple code versions, each being optimized for specific hardware features. The oneAPI framework, recently introduced by Intel, contains a set of programming tools for the development of portable codes that can be compiled and fine-tuned for CPUs, GPUs, FPGAs, and accelerators. In this paper, we report on the experience of porting the implementation of Boris particle pusher to oneAPI. Boris particle pusher is one of the most demanding computational stages of the Particle-in-Cell method, which, in particular, is used for supercomputer simulations of laser-plasma interactions. We show how to adapt the C++ implementation of the particle push algorithm from the Hi-Chi project to the DPC++ programming language and report the performance of the code on high-end Intel CPUs (Xeon Platinum 8260L) and Intel GPUs (P630 and Iris Xe Max). It turned out that our C++ code can be easily ported to DPC++. We found that on CPUs the resulting DPC++ code is only ~10% on average inferior to the optimized C++ code. Moreover, the code is compiled and run on new Intel GPUs without any specific optimizations and shows the expected performance, taking into account the parameters of the hardware.

## 1    Introduction

The development of computational architectures in the last decades has led to the emergence of new possibilities for supercomputer simulations. However, the appearance of devices with fundamentally different architectures required the development of appropriate approaches to programming and code optimization. It turned out that the development of a universal framework that allows implementing a single code that can be compiled and, no less important, work efficiently on different hardware is not straightforward. Such frameworks and libraries, in particular, include OpenCL [1], OpenACC [2], Alpaka [3], Kokkos [4], and many others. In 2020, Intel introduced oneAPI – a new unified open model for heterogeneous programming, which includes a wide set of tools and a new DPC++ language [5] for heterogeneous programming based on the SYCL language. The DPC++ language allows using various computing devices in calculations, in particular, CPUs, GPUs, FPGAs, and other accelerators.

[*] This study is supported by Intel Corporation (oneAPI Center of Excellence program) and by the Ministry of Science and Higher Education of the Russian Federation, project no. 0729-2020-0055.

In this paper, we report on the experience of porting the algorithm of Boris pusher to the DPC++ programming language. The Boris pusher is a frequently used algorithm for advancing the classical state (coordinate and momentum) of a charged particle under the action of a given electromagnetic field. This algorithm is one of the main computational cores of the High-Intensity Collisions and Interactions (Hi-Chi) framework [6, 7], which is an open-source collection of Python-controlled tools for performing simulations and data analysis in the research area of strong-field particle and plasma physics. In particular, we address the following questions. Firstly, we demonstrate how such code can be ported to DPC++. Secondly, we analyze the performance of the DPC++ code on high-end Intel CPUs versus the baseline C++ implementation and show how the key code optimization techniques affect performance in different simulation scenarios. Finally, we assess the performance of the DPC++ code on new Intel GPUs versus CPUs without any additional optimizations for GPUs.

The paper is organized as follows. Section 2 provides a general overview of the Particle-in-Cell method and, in particular, the particle push algorithm. In Section 3 we describe main data structures and algorithms. In Section 4 we propose the new parallel implementation of the particle pusher based on the recently introduced DPC++ programming language. Section 5 presents numerical results and discussion. Section 6 concludes the paper.

## 2     Method

In this subsection we briefly describe the Particle-in-Cell method; a detailed description is given in [8]. The Particle-in-Cell method is used to model the interaction of an electromagnetic field with plasma using kinetic description. This method operates on two distinct sets of data: grid field data and particle data. The values of electric and magnetic fields are defined on a spatial grid. The plasma is represented as an ensemble of particles, each with a charge, mass, position and momentum. Each particle used in simulation is in fact a macroparticle that represents a cloud of real particles, whose distribution is described by a fixed localized shape function, also referred to as the form factor of a macroparticle. A notable feature of the method is that particles do not interact with each other directly; instead each particle interacts with a set of nearby grid values of the electromagnetic field, depending on the form factor.

The conventional computational loop of the Particle-in-Cell method consists of four stages. Field values are updated by solving Maxwell's equations

$$\frac{\partial \boldsymbol{E}}{\partial t} = c\, rot\boldsymbol{B} - 4\pi \boldsymbol{J}, \quad (1)$$

$$\frac{\partial \boldsymbol{B}}{\partial t} = -c\, rot\boldsymbol{E}, \quad (2)$$

where $\boldsymbol{E}$ and $\boldsymbol{B}$ are electric and magnetic fields, respectively, $\boldsymbol{J}$ is the electric current density produced by particle motion, $c$ is the speed of light. These equations can be solved using FDTD [9] or FFT-based [8] techniques. For each particle the Lorenz force is computed using interpolated values of the electromagnetic field and the particle momentum and position are updated. The grid values of the current $\boldsymbol{J}$ are computed and added to Maxwell's equations forming the self-consistent system of equations.

This article concerns one of the main parts of the Particle-in-Cell method: the integration of particle motion in electromagnetic fields. This stage, usually called the Particle push, is of particular interest for performance optimization, because this stage becomes the most time consuming for realistic problems due to a large number of macroparticles (as compared to the number of grid nodes). At this stage the equations of motion are solved together with Newton's second law accounting for relativistic effects. Usually in this case the consideration is restricted to the Lorentz force. The system of equations can be written as

$$\frac{d\boldsymbol{r}}{dt} = \boldsymbol{v}, \quad (3)$$

$$\frac{d\boldsymbol{p}}{dt} = q\left(\boldsymbol{E} + \frac{1}{c}\boldsymbol{v}\times\boldsymbol{B}\right), \quad (4)$$

$$\boldsymbol{p} = \frac{m\boldsymbol{v}}{\sqrt{1-\frac{v^2}{c^2}}} = \gamma m\boldsymbol{v}, \quad (5)$$

where $\boldsymbol{r}, \boldsymbol{v}, \boldsymbol{p}$ are position, velocity and momentum, $m, q$ are charge and rest mass, $\gamma$ is the Lorenz factor of macroparticle, respectively. This set of equations allows different numerical scheme for integration both explicit and implicit [10]. New methods were recently presented and their comprehensive study was given in Ref. [11]. Nevertheless the most used and *de-facto* standard scheme is the Boris method [12]. The numerical code Hi-Chi also uses this method. This method is described in detail in [8], but for the clarity of presentation a short description is given here.

For efficient implementation velocity and position are displaced by half a time step and their integration leap over each other. This scheme allows avoiding problems related to numerical acceleration of particles moving in a magnetic field. The finite-difference approximation of equations (3) and (4) can be written as

$$\frac{\boldsymbol{p}^{n+\frac{1}{2}} - \boldsymbol{p}^{n-\frac{1}{2}}}{\Delta t} = q\left(\boldsymbol{E}^n + \frac{1}{c}\bar{\boldsymbol{v}}\times\boldsymbol{B}^n\right), \quad (6)$$

$$\frac{\boldsymbol{r}^{n+1} - \boldsymbol{r}^n}{\Delta t} = \boldsymbol{v}^{n+\frac{1}{2}}, \tag{7}$$

where $\bar{\boldsymbol{v}}$ is the velocity averaged over the time step, superscript $n$ refers to time moment $t_n = n\Delta t$, and $\Delta t$ is the numerical scheme time step. Here we assume that the particle momentum $\boldsymbol{p}^{n-\frac{1}{2}}$ and the particle position $\boldsymbol{r}^n$ are known. We need to advance them to $\boldsymbol{p}^{n+\frac{1}{2}}$ and $\boldsymbol{r}^{n+1}$, respectively. Note that the states of electric and magnetic fields can also be shifted in time by a half step if the FDTD is used, but we disregard this aspect in our further consideration. The expression for Lorentz force (6) includes the average velocity $\bar{\boldsymbol{v}}$. The correct choice of expression for $\bar{\boldsymbol{v}}$ is not straightforward in relativistic case and may affect the results of simulations, see Ref. [11] for details. In the present paper we address the conventional Boris method, for which the expression for average velocity is written as

$$\bar{\boldsymbol{v}} = \frac{\boldsymbol{p}^{n+\frac{1}{2}} + \boldsymbol{p}^{n-\frac{1}{2}}}{2\gamma^n m}. \tag{8}$$

The elegant idea of integrating this equation was to split the step into a symmetrized sequence of two half steps due to electric field with a full step due to magnetic field in between. This can be realized by substitutions

$$\boldsymbol{p}^{n-\frac{1}{2}} = \boldsymbol{p}^- - q\boldsymbol{E}\frac{\Delta t}{2}, \tag{9}$$

$$\boldsymbol{p}^{n+\frac{1}{2}} = \boldsymbol{p}^+ + q\boldsymbol{E}\frac{\Delta t}{2}, \tag{10}$$

that lead to the equation

$$\frac{\boldsymbol{p}^+ - \boldsymbol{p}^-}{\Delta t} = \frac{q}{2c\gamma^n m}(\boldsymbol{p}^+ + \boldsymbol{p}^-) \times \boldsymbol{B}. \tag{11}$$

By considering a scalar multiplication of both sides by $(\boldsymbol{p}^+ + \boldsymbol{p}^-)$ we can see that the equation preserves $p^2$ (i.e. $(\boldsymbol{p}^+)^2 = (\boldsymbol{p}^-)^2$). From the limit of small $\Delta t$ we see that the equation describes a pure rotation of vector $\boldsymbol{p}$ about vector $\boldsymbol{B}$ to an angle $\Delta t q (2c\gamma^n m)^{-1}$, where $\gamma^n = \gamma(\boldsymbol{p}^+) = \gamma(\boldsymbol{p}^-) = (1 + p^2/(m^2 c^2))^{1/2}$. Under the assumption that the rotation angle is small (i.e. the time step is sufficiently small) the advanced state $\boldsymbol{p}^+$ can be approximately obtained without computing trigonometric functions so that $p^2$ is preserved exactly (i.e. independently of the smallness of the rotation angle):

$$\boldsymbol{p}' = \boldsymbol{p}^- + \boldsymbol{p}^- \times \boldsymbol{t}, \boldsymbol{p}^+ = \boldsymbol{p}^- + \boldsymbol{p}' \times \boldsymbol{s}, \tag{12}$$

where auxiliary vectors (see [8] for more the details):

$$\boldsymbol{t} = \frac{q\boldsymbol{B}}{\gamma^n mc}\frac{\Delta t}{2}, \quad \boldsymbol{s} = \frac{2\boldsymbol{t}}{1+t^2}. \tag{13}$$

Summarizing, in this method the particle motion is performed in a following procedure:

1. Perform half-step due to $E$ to obtain $p^-$ from $p^{n-\frac{1}{2}}$ using equation (9).
2. Perform full rotation using equation (12) to find $p^+$ using auxiliary vectors $s$ and $t$ defined in (13).
3. Perform half-step due to $E$ to obtain $p^{n+\frac{1}{2}}$ from $p^+$ using equation (10).
4. Calculate $v^{n+\frac{1}{2}}$ based on $p^{n+\frac{1}{2}}$ and advance particle position to $r^{n+1}$ according to (7).

The problem of high-performance parallel implementation of the Particle-in-Cell method is well studied [13-19]. In this paper we discuss the Boris pusher, one the main computational kernels of many such codes, and its implementation developed on C++ as a part of the Hi-Chi numerical code. Porting and optimization of the Boris method using the DPC++ language is discussed further in this paper.

## 3   Data Structures and Algorithm

The developments reported in this paper are a part of the Hi-Chi project [6]. The project Hi-Chi is an open-source collection of Python-controlled tools for performing simulations and data analysis in the research area of strong-field particle and plasma physics. The tools are being developed in C++ and provide high performance using either local or supercomputer resources. The project is intended to offer an environment for testing, benchmarking and aggregative use of individual components, ranging from basic routines to supercomputer codes.

A `Particle` class is the key data structure used in our simulations. For each particle, we store position and momentum vectors of 3 floating point numbers each, as well as scalar floating point values of the particle weight and the Lorenz factor γ. Additionally, we store an integer value of the particle type to determine its mass and charge. These parameters, corresponding to particles of different types, are stored in a separate table in a single copy. Thus, the data in the Particle class is described as follows (see definition of FP and FP3 below):

```
Class Particle {
    FP3 position; // Particle position (x, y, z)
    FP3 momentum; // Particle momentum (px, py, pz)
    FP weight;    // Particle weight
    FP gamma;     // Particle γ-factor
    Short type;   // Particle type
    …
};
```

The code is implemented so that we can easily switch between using single and double precision data types. To do this, we abstracted the floating point data type as `FP`, which can be `float` or `double` depending on the settings. Similarly, the `FP3` data type describes a vector of 3 `float` or `double` components. In the case of single precision, storage of each particle requires 34 bytes of memory (36 bytes after memory alignment), in the case of double precision, each particle takes 66 bytes of memory (72 bytes after memory alignment). The investigation of the possibility of performing calculations in single and double precision is beyond the scope of this study. Here we are only comparing the performance of calculations in single and double precision. We should also note that in the considered benchmarks, we did not observe any inaccuracies caused by the use of single precision.

The way of organizing an ensemble of particles deserves special attention. For example, in programs for supercomputer modeling of laser plasma by the particle-in-cell method, two main approaches of representing an ensemble of particles are commonly used. The first method assumes that each cell stores its own array of particles. This representation has many advantages, but it requires handling the movement of particles between cells, which causes an additional overhead when parallelizing computations. The second way is to store the entire ensemble of particles in a single array. In this case, we do not need to handle the movement of particles between cells, but we have to periodically sort the array of particles in order to improve cache locality. In the Hi-Chi code, we employ the second method.

The next question that arises when choosing data structures to represent an ensemble of particles is which of the common patterns is better to use: an *array of structures* (AoS, in our case, an array of objects) or a *structure of arrays* (SoA). This issue has been studied for a long time as applied to various problems. It is known that both approaches of data representation have their pro et contra. For example, the AoS pattern allows us to preserve memory locality. However, this scheme is not very efficient in the case of code vectorization, since it entails non unit-stride access to the data of different particles. On the contrary, the SoA pattern is less efficient in utilizing cache memory, but it allows us to efficiently load data for vector computations and does not use time consuming scatter/gather operations. In the general case, none of the schemes is unconditionally better. Everything is determined by the properties of the algorithm, problem, and target architecture. Therefore, Hi-Chi allows one to use any of these patterns. Next, we will compare how the choice of data structures affects the performance of the code.

Note that in order to use different ways of storing data, we implement the `ParticleProxy` class, which completely repeats the functionality of the `Particle` class, but stores *references* to objects. This approach allows us to effectively employ the C++ templates and use the single code regardless of the storage structure.

## 4   Exploiting Parallelism Using the oneAPI Technology

### 4.1   Reference Implementation of the Boris Pusher

As a reference implementation, we consider a parallel version implemented using the OpenMP technology. Parallelism in this version is exploited at the level of particle processing, and the loop over particles is *parallelized* and *vectorized* as follows:

```
// Numerical integration loop over numSteps time steps
for (int step = 0; step < numSteps; step++)  {
    // Run the Pusher for every particle in an ensemble
    #pragma omp parallel for simd
    for (int ind = 0; ind < numParticles; ind++)  {
        // Run the Boris pusher for particle #ind
        …
    }
}
```

### 4.2   Porting the Pusher to DPC++

Smart memory management is a key factor to achieving good performance and scalability of codes. In the case of using accelerators, this issue becomes even more important. DPC++ provides two ways to manage memory and access/share/move data between devices. The first method involves the use of special concepts – buffers, which allow us to define regions of memory that can be used on the device (buffers), and accessors, which allow us to plan access to data and their movement between devices. The second method (Unified Shared Memory, USM) is more low-level and allows us to work in a style similar to working with C++ pointers. This model is quite convenient for codes that have been already based on C++ pointers. In this case, porting to DPC++ requires just minimal modifications to memory allocation instructions.

We employ the USM model. It is the simplest, but quite functional option for shared memory allocation providing data access on a device and a host. We also rely on oneAPI runtime for memory management. This approach allowed us to quickly port the code to DPC++, with only minimal changes and reasonable performance. Compared to the reference implementation, our DPC++ code is quite similar:

```
// Numerical integration loop over numSteps time steps
for (int step = 0; step < numSteps; step++) {
    // Create a "kernel" function
    auto kernel = [&](sycl::handler& h) {
        // Work with particles in parallel
        h.parallel_for(sycl::range<1>(numParticles),
                    [=](sycl::id<1> ind) {
            // Run the Boris pusher for particle #ind
            …
        }
    }
    // Submit the kernel
    device.submit(kernel).wait_and_throw();
}
```

The code, as before, processes the movement of particles in parallel. Unlike typical C++ code, for processing particles, we create a kernel using special C++ lambda expression (supported since the C++ 11 standard). This kernel employs a special

DPC++ mechanism `parallel_for`, which calls the Boris pusher in parallel for particles from the ensemble. Code vectorization is also automatically provided by the compiler. Since the Boris pusher is implemented as a lambda expression that captures objects *by copy*, these objects must have a default copy constructor that will create full copies of objects with the same addresses in memory. Therefore, we could not use the standard vector class to implement an array of particles. Instead, we use a C-style pointer to a buffer, which is copied without actually copying the contents of the buffer when capturing objects to the kernel. Such copying is usually a mistake for C++ classes, but in this case it is exactly the required behavior.

### 4.3 Improving Scaling Efficiency

DPC++ runtime on a CPU employs the widely used Threading Building Blocks (TBB) library for parallel computations. Compared to OpenMP, TBB always uses dynamic scheduling, which can substantially improve performance in complex unbalanced problems. However, in balanced applications, the overhead of dynamic scheduling may not be justified. However, a small overhead is a reasonable price to pay for the versatility of the code that can be compiled and run on different architectures.

Appropriate use of platforms with Non-Uniform Memory Access (NUMA) architecture deserves a separate discussion. Thus, on modern supercomputers, a configuration with several (often two) CPUs is typical. In such cases the access of the cores to the local memory of their processor is much faster than access to the memory of another processor installed on the same node. This is especially important for memory-bound applications, in particular for the considered pusher.

In codes parallelized with OpenMP, we can often achieve that the data is localized in the cache memory of the CPU that will process it. In the case of using TBB (recall that DPC++ uses this scenario), we can also work with memory in a *NUMA-friendly* manner. In this regard we use the `DPCPP_CPU_PLACES` environment variable with the value `numa_domains`. In this case, the iteration space is divided into NUMA domains, and TBB performs dynamic scheduling of parallel execution of tasks only within the corresponding NUMA arena. This ensures that the same particles are processed on the same CPU at every time step. It will be shown below that this significantly improves performance and scaling efficiency of the code. In what follows, we will refer to such launches as 'DPC ++ (NUMA)'.

## 5 Numerical Results

### 5.1 Computational Infrastructure

The computational experiments were performed at a node of the supercomputer Endeavour with 2x Intel Xeon Platinum 8260L (Caskade Lake, 24 cores each), 48 cores overall, 192GB RAM, RedHat 4.8.5, Intel C++ Compiler and Intel DPC++ Compiler from the Intel OneAPI Toolkit Base and HPC (Gold Release 2020) suite. All tests on Intel P630 and Iris Xe Max GPUs were executed on Intel DevCloud.

Some preliminary tests were executed on the Lobachevsky supercomputer at Lobachevsky University. The hardware parameters are presented in Table 1.

**Table 1.** Hardware parameters

| Parameter | 2x Intel Xeon Platinum 8260L | P630 | Iris Xe Max |
|---|---|---|---|
| Number of CPU cores / GPU execution units | 48 | 24 | 96 |
| Clock frequency | 2.4 GHz (3.9 GHz Boost) | 0.35 GHz (1.15 GHz Boost) | 0.3 GHz (1.65 GHz Boost) |
| RAM | DDR4 192GB | DDR4 32 GB (CPU RAM) | LPDDR4X 4 GB |
| Peak performance (single precision) | 3.6 TFlops | 0.441 TFlops | 2.5 TFlops |

### 5.2 Benchmarks

We considered *two simulation scenarios* as benchmarks for analyzing performance. In the *first scenario*, all field values are precalculated and stored in the corresponding array. This scenario allows excluding all operations from measurements except for particle motion. The *second scenario* assumes that the fields are specified analytically. In this case, we do not have to store a large data array. On the contrary, field values are computed using analytical formulas when they are directly needed in calculations. Both scenarios are in demand in practice and, hypothetically, can lead to different conclusions regarding code optimization, since in the first case, we store much more data, and in the second, we perform much more calculations.

In order to test our implementations we consider the motion of electrons in the tightly focused fields in the form of a standing magnetic dipole (m-dipole) wave [20]. This study is necessary to determine the optimal parameters of a seed target for the vacuum breakdown in multipetawatt m-dipole wave [21]. Tight focusing allows decreasing of the threshold power of this phenomenon [22] that is favourable for upcoming experiments at 10-PW laser facilities [23]. For this reason, we consider ultimate focusing [24] in a form of the dipole wave.

The pulsed multi-PW incoming m-dipole wave can ionize matter at its leading edge and pull unbound electrons to the wave focus. When the wave passes through the focus the diverging wave appears and electrons start to oscillate in the standing wave. In order to trigger the vacuum breakdown a number of particles should remain in the focus when the instantaneous wave power becomes greater than 10 PW [21]. However, due to strong field inhomogeneity, particles can rapidly escape the focal region while instantaneous power is not high enough. With the help of simulations of the particle motion in the standing m-dipole wave the rate of particle escape from the focal region can be obtained. Based on these results the optimal parameters of the seed target can be chosen.

Particle escape is fastest in the range of powers from approximately 4 GW to 1 PW when fields are relativistic, but radiative trapping effects [25] are absent. For the test we consider the wave power P = 0.1 PW. In the simulation the electric and magnetic field components are set analytically as follows:

$$E_x = -\frac{2A_0 y}{R(x,y,z)} \cos(\omega_0 t) f_1(R(x,y,z))$$
$$E_y = \frac{2A_0 x}{R(x,y,z)} \cos(\omega_0 t) f_1(R(x,y,z))$$
$$E_z = 0$$
$$B_x = -\frac{2A_0 xz}{R^2(x,y,z)} \sin(\omega_0 t) f_2(R(x,y,z))$$
$$B_y = -\frac{2A_0 xy}{R^2(x,y,z)} \sin(\omega_0 t) f_2(R(x,y,z))$$
$$B_z = -\frac{2A_0 z^2}{R^2(x,y,z)} \sin(\omega_0 t) \left(\frac{z^2}{R^2(x,y,z)} f_2(R(x,y,z)) + f_3(R(x,y,z))\right),$$

(14)

where $t$ is time, $x, y, z$ are Cartesian coordinates, $R(x,y,z) = \sqrt{x^2 + y^2 + z^2}$, $A_0 = k\sqrt{3P/c}$, $c$ is the light velocity, $\omega_0 = 2.1 \times 10^{15} \text{s}^{-1}$ is the wave frequency corresponding to the wavelength $\lambda = 0.9\mu\text{m}$, $k = \omega_0/c$,

$$f_1(R) = \frac{\sin(kR)}{(kR)^2} - \frac{\cos(kR)}{kR}$$
$$f_2(R) = \left(\frac{3}{(kR)^3} - \frac{1}{kR}\right) \sin(kR) - \frac{3\cos(kR)}{(kR)^2}$$
$$f_2(R) = \left(\frac{1}{kR} - \frac{1}{(kR)^3}\right) \sin(kR) + \frac{\cos(kR)}{(kR)^2}.$$

(15)

Initially ($t = 0$), electrons are at rest and distributed uniformly within the sphere with radius $r = 0.6\lambda$. The experimental setup is as follows. In each experiment, $10^7$ particles were simulated. The equations of motion were integrated over $10^3$ time steps, which we further refer to as *'iteration'*. During the experiment, 10 successive iterations were measured. To compare the performance results, we used the **NSPS** metric (*nanoseconds per particle per step*) calculated as the average time of one iteration in nanoseconds, divided by the number of particles ($10^7$) and by the number of steps in one iteration ($10^3$).

### 5.3    Results and Discussion

**Experiments on CPUs**
First of all, it is necessary to take into account the following fact. When profiling computational codes, we often observe that the first iteration of a method can take work slower than the rest. This is usually explained by the fact that at the first iteration, the data has to be loaded from RAM, while at the next iterations, part of the data is loaded from a cache. In NUMA systems, this effect is sometimes even more pronounced if the code does not implement a NUMA-friendly memory usage policy. In

4the case of DPC++ codes, this effect is manifested in an even more explicit form, since when the kernel is first launched, it is compiled from an intermediate representation for a specific hardware, which can take some time. In our benchmark, the first iteration takes 50% longer time than the subsequent ones, which is the cumulative effect of the reasons described above. Considering that we perform a lot of iterations, this effect does not have a significant impact on the results.

We collected the results on CPUs employing available 48 cores (2 CPUs with 24 cores each). The comparison involves implementations parallelized on OpenMP, or DPC++, or DPC++ with the NUMA-friendly memory usage policy described before. For each of these implementations, we tried using SoA and AoS memory layout patterns. As stated earlier, two simulation scenarios were considered. We refer them to as 'Precalculated Fields' and 'Analytical Fields'. For OpenMP versions, it was found that employing 96 threads is empirically the best, that is, the use of hyperthreading technology improves performance. For DPC++ implementations, the number of threads is selected by the TBB runtime. All experiments were executed both in single and in double precision (Table 2).

**Table 2.** Performance results (NSPS, nanoseconds per particle per step) on CPU for 6 implementations and 2 simulation scenarios.

| Pattern | Parallelization | Precalculated Fields | | Analytical Fields | |
|---|---|---|---|---|---|
| | | float | double | float | double |
| AoS | OpenMP | 0.53 | 0.98 | 0.58 | 0.84 |
| | DPC++ | 0.78 | 1.54 | 1.02 | 1.48 |
| | DPC++ NUMA | 0.54 | 0.99 | 0.54 | 0.89 |
| SoA | OpenMP | 0.50 | 1.06 | 0.43 | 0.76 |
| | DPC++ | 0.85 | 1.49 | 0.77 | 1.31 |
| | DPC++ NUMA | 0.58 | 1.20 | 0.60 | 0.90 |

The results lead to the following conclusions:

1. Using the NUMA-friendly memory usage policy leads to a significant performance gain due to the elimination of the overhead of remote access to the memory of another CPU installed on the same node. Note that in the OpenMP code, similar tricks did not lead us to performance improvement, since in this case remote access occurs only at the first time steps of the method, then the data is localized within the corresponding NUMA domains. The conclusions are confirmed by profiling using Intel VTune. Note also that although such a significant effect of NUMA on performance is specific to the considered memory bound benchmark, it can be important for optimizing other DPC ++ applications as well.
2. The performance of the optimized DPC++ implementation is only slightly inferior to the OpenMP implementation. The difference is usually only ~10% on average due to some overhead and a different approach to parallelization. We think this is an excellent result for DPC++ taking into account the portability of the code.

3. The choice of the AoS or SoA patterns has almost no effect on the performance in the current benchmark. This is due to the fact that the main factor limiting performance is not loading data into vector registers, but working with RAM.
4. When going from single to double precision, the running time changes as expected, because it requires twice the memory bandwidth and doubles the amount of computation. In the problem with precomputed fields, the difference is almost twofold; in the case of analytical fields, it is slightly less due to the specifics of the calculations. Note also that in the case of DPC++, code vectorization occurs with full use of AVX-512 instructions, as it was earlier in OpenMP.
5. Since the problem is memory bound, working with memory dramatically affects performance. The two considered simulation scenarios are fundamentally different in working with memory, since in the 'Precalculated Fields' problem, we additionally store an array of field values comparable in size to the ensemble of particles. On the contrary, in the 'Analytical Fields' problem, we do a lot more resource-intensive calculations of mathematical functions. The main motivation for considering these two scenarios was to find out how these differences affect the overall simulation time. The results showed that calculations using analytical formulas and loading pre-calculated data from memory turned out to be, on the whole, comparable in terms of time consumption. At the same time, in the case of calculations in double precision, the scenario with the analytical computations of field values runs a little faster. It is noteworthy that this result does not depend on the choice of parallel programming technologies (OpenMP or DPC++).

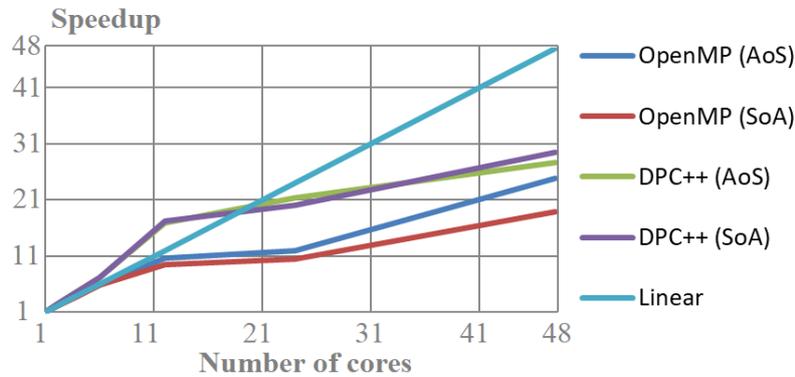

**Fig. 1.** Speedup of parallel computations of the OpenMP and DPC++ NUMA implementations employing the AoS and SoA data layouts in the 'Precalculated Fields' problem. Computations are performed in single precision on 1–48 cores. Single core run time is used as a reference.

To evaluate the efficiency of parallelization, we calculate the speedup when using 1–48 cores relative to runs on a single core. Considering that hyperthreading is enabled, we start 2 threads on each core, binding threads to cores. As an example, single precision calculations in the problem with precalculated fields are considered. The results (Fig. 1) show that in the implementation on OpenMP, a close to linear speedup is observed until the code fully utilizes memory bandwidth of the first socket. When

13we start using of the second socket, the run time begins to scale linearly again. For DPC++ NUMA implementations, super-linear acceleration is observed at the beginning. This is because the DPC++ single core version is quite slow. Further experiments demonstrate reasonable scaling, approaching to 63% of strong scaling efficiency when using 48 cores. As shown earlier, the overall run times for OpenMP and DPC++ NUMA versions are close to each other.

**Experiments on GPUs**

DPC++ is the universal development tool for portable programs, which is a great achievement for the development team. However, achieving *performance portability* is even much more complex problem due to fundamental differences in computing architectures. Apparently, when porting DPC++ codes to specific architectures, some fine-tuning or even new implementations of the computational kernels can be required. One of the goals of the present work was to study how the DPC++ code, built on the basis of the C++ code optimized for Intel CPUs, will work on the new Intel GPUs without any specific optimizations. The results obtained should not be taken as a fair comparison of CPUs vs. GPUs, they only demonstrate how much performance we can get without additional work. We carried out such experiments on Intel devCloud, using currently available devices, the parameters of which are shown earlier in Table 1. Since for the Iris Xe Max, double precision operations occur only in an emulation mode, we present the results in single precision only. The results are shown in Table 3.

**Table 3.** Performance results (NSPS, nanoseconds per particle per step) on GPUs for DPC++ implementations in 2 simulation scenarios. Computations are performed in single precision.

| Pattern | Precalculated Fields | | | Analytical Fields | | |
|---|---|---|---|---|---|---|
| | CPU | P630 | Iris Xe Max | CPU | P630 | Iris Xe Max |
| AoS | 0.54 | 4.76 | 2.10 | 0.54 | 4.45 | 2.10 |
| SoA | 0.58 | 2.43 | 1.42 | 0.60 | 1.93 | 1.00 |

If for the CPUs different particles memory layouts were comparable in performance due to various factors described earlier, then on Intel GPUs the run time may differ by more than half (Table 3). This is due to a different organization of the memory subsystem in the GPUs. We should also note the lack of additional optimizations for the GPUs. Probably, the performance of the AoS version of the code can be improved, however, in any case, the importance of choosing a layout on GPUs must be taken into account when such porting. A direct comparison of the run time on the CPUs and GPUs is also of great interest. As stated earlier, this comparison is not fully objective due to the lack of GPU optimizations. However, it provides an answer to the question of whether we can expect the GPU run time to be appropriate after such porting. In the problems we are considering, we can give a positive answer to this question. Indeed, as compared to the considered Xeon CPUs, the performance of P630 and Iris Xe Max is lower by a factor of about 8 and 1.5, respectively. At the same time, the code on P630 works slower only by a factor of 3.5–4.5, and the code on Iris Xe Max is slower by a factor of 1.7–2.6, compared to 2 high-end CPUs. This comparison does not give a complete picture, since GPUs have a different memory organization, the problem is not compute- but memory-bound, and utilization of GPUs is often much harder compared to CPUs. Nevertheless, we can conclude that even without additional optimizations, we got reasonable performance on GPUs, which, most likely, can be further improved.

## 6       Conclusion

The paper presents a new DPC++ implementation of the Boris Pusher algorithm for the movement of particles in a given electromagnetic field. The implementation is obtained by porting the CPU-optimized C++ implementation in the Hi-Chi code by replacing the way of organizing parallel computations. It turned out that this porting can be done quickly enough. After running the program on the high-performance server with 2 high-end CPUs, we found that the performance of the resulting DPC++ implementation significantly depends on the run settings customization in terms of optimal use of the NUMA architecture, while the SoA and AoS patterns of the data layout have almost no effect on performance. As a result, it was found that, regardless of the simulation scenario, the DPC++ implementation is only slightly inferior to the C++ code, while it became possible to run it on Intel GPUs.

Our experiments on Intel GPUs showed that even though we did not optimize the code for the GPU in any way, the performance results compared to the optimized code on the CPUs exceed our expectations. So, it turned out that 2 Xeon CPUs are ahead of desktop GPUs only in accordance with the difference in peak performance capabilities. We expect that the performance of the GPU implementation can be improved. This is one of the directions for further research.

The code is publicly available [6].